# Entanglement Entropy Fluctuations in Quantum Ising Chains

## M. A. Yurishchev

*Institute of Problems of Chemical Physics, Russian Academy of Sciences, Chernogolovka, Moscow oblast, 142432 Russia*
*e-mail: yur@itp.ac.ru*

**Abstract**—The behavior of Ising chains with the spin–spin interaction value λ in a transverse magnetic field of constant intensity ($h = 1$) is considered. For a chain of infinite length, exact analytical formulas are obtained for the second central moment (dispersion) of the entropy operator $\hat{S} = -\ln\rho$ with reduced density matrix ρ, which corresponds to a semi-infinite part of the model chain occurring in the ground state. In the vicinity of a critical point $\lambda_c = 1$, the entanglement entropy fluctuation $\Delta S$ (defined as the square root of dispersion) diverges as $\Delta S \sim [\ln(1/|1-\lambda|)]^{1/2}$. For the known behavior of the entanglement entropy $S$, this divergence results in that the relative fluctuation $\delta S = \Delta S/S$ vanishes at the critical point, that is, a state with almost nonfluctuating entanglement is attained.

## 1. INTRODUCTION

Quantum entanglement is a key concept in quantum informatics, cryptography, and teleportation. In recent years, this concept has also occupied an important place in the quantum field theory, the black-hole problem, theory of phase transitions, etc. There are several approaches to the quantitative characterization of entanglement. The most promising approach is based on using the entropy of a reduced density matrix. The so-called entanglement entropy can be interpreted as the amount of information that is accessible to an observer, for which some of the variables (degrees of freedom) of the complete system during measurements is lost (or inaccessible). For a more detailed consideration of the issue of entanglement see, e.g., [1–8] (and numerous references therein).

The density matrices (as statistical operators) were originally introduced into science in 1927 by Landau [9] for the description of the decay phenomena in wave mechanics and then independently and more systematically presented by von Neumann [10, 11]. In 1930, the simple reduced (one-particle) density matrices were explicitly employed in the quantum theory by Fock [12] and somewhat later by Dirac [13] for developing the methods of Hartree and Thomas, respectively.

The idea of excluding excess information by means of integration (summation, convolution) over most variables in a density matrix (or a distribution function) of a multiparticle system and the derivation of equations for the reduced density matrix of a lower order was later used by many researchers [14–16]. An important achievement was the introduction of a hierarchical chain of coupled Bogolyubov–Born–Green–Kirkwiod–Yvon (BBGKY) equations [17, 18]. The truncation of the infinite chain and its closure within a certain approximation allowed a number of problems for moderate-density media to be solved.

Another direction of application of the reduced density matrix was found in 1986, when Bombelli et al. [19] successfully used it to calculate the entropy for a black hole model, which made it possible to explain the Bekenstein–Hawking law of areas, according to which black-hole entropy is proportional to the area of its horizon of events rather than to the volume [20, 21]. Then, the notions of entanglement entropy were used in the gauge theories and string theory [22–25].

In 1996, the entropy of a reduced density matrix was imparted a new important information meaning; it was found to be equal to a relative number of maximum entangled pairs, which could be extracted from a large number of copies of the initial system using a purification protocol involving only local operations and classical communications (LOCC) [26, 27]. Since then, the entanglement of a two-component system as identified with the entropy of a density matrix has been widely used in the quantum information physics.

However, as is well known from statistical physics, entropy is subject to fluctuations. Therefore, generally speaking, one must also admit the fluctuations of entanglement [28–30]. Indeed, the first calculations for two-qubit systems already showed that these fluctuations can be large [28]. The fluctuations arise due to a reservoir that appears for a given subsystem upon averaging over the degrees of freedom of the remaining part of the system.

The present investigation is devoted to spin-1/2 quantum Ising chain models on one-dimensional lattices with free boundary conditions. The chains are divided into two equal parts (subsystems), one of which is subjected to reduction. The analysis starts with a dimer, followed by a transition to a chain of infinite length.

In the limit of an infinitely long Ising chain and, accordingly, semi-infinite subsystems, the entropy of entanglement $S$ was calculated by Calabrese and Cardy [31, 32]. In the present study, a dispersion of the entanglement entropy for this system is calculated. It is found that this quantity, as well as the entanglement itself, exhibits logarithmic divergence at the phase transition point. It will be also demonstrated that fluctuations $\Delta S$ of the entanglement entropy in an ordered (ferromagnetic) phase are always smaller than the $S$ value proper. On the contrary, 99.995% of a disordered (paramagnetic) region with respect to the spin–spin interaction magnitude $\lambda$ corresponds to strong fluctuations such that $\Delta S > S$.

The Introduction is followed by Section 2, which gives quantum-mechanical definitions of the density matrix, entanglement entropy, and the magnitude of its fluctuations. Section 3 formulates the model and considers a system of two qubits. Section 4 is devoted to deriving formulas for the entanglement entropy fluctuations for the Ising chain in the thermodynamic limit. The obtained results are thoroughly considered in Section 5 and briefly summarized in the final Section 6. The appendix provides useful relationships from the theory of elliptic functions, which allowed the formulas for fluctuations of the entropy of entanglement to be presented in quadratures (via complete elliptic integrals of the first and second kinds).

## 2. DEFINITIONS OF $S$ AND $\Delta S$

Following the commonly accepted courses of quantum mechanics and statistical physics (see, e.g., [33–35]), consider system $X$ embedded into (surrounded by) another system $Q$ (reservoir) so that the total system $X \cup Q$ is closed. If $|\psi\rangle$ is the wave function (vector) of the total system, the state of system $X$ is described by the following density matrix:

$$\rho = \mathrm{Tr}_Q |\psi\rangle\langle\psi|, \qquad (1)$$

where the trace is taken over all degrees of freedom of reservoir $Q$. The dimensionless entropy operator of the system under consideration is as follows:

$$\hat{S} = -\ln\rho \qquad (2)$$

and the entropy of system $X$ is

$$S = \mathrm{Tr}(\rho\hat{S}) = -\mathrm{Tr}(\rho\ln\rho). \qquad (3)$$

According to [26, 27], this quantity represents the entropy of entanglement (measured in nats).

Relation (3) yields an average value or the first moment of the entropy operator (2). Then, the reduced second-order moment yields the dispersion of entropy as

$$D = \mathrm{Tr}[\rho(\hat{S}-S)^2] = \mathrm{Tr}[\rho(\hat{S})^2] - [\mathrm{Tr}(\rho\hat{S})]^2. \qquad (4)$$

In terms of statistical mechanics, the square root of this dispersion determines the root-mean-square fluctuation as follows:

$$\Delta S = \sqrt{D} = [\langle(\hat{S}-\langle\hat{S}\rangle)^2\rangle]^{1/2}, \qquad (5)$$

where angular brackets denote averaging over an ensemble with the density matrix $\rho$.

Note that, according to formula (1), all density matrices are, in fact, reduced except for $|\psi\rangle\langle\psi|$ for a closed system. The usual statistical entropy (3) in the quantum theory of information is used as a measure of entanglement. However, this choice of measurements has a disadvantage in that the entanglement entropy is subject to fluctuations, as a result of which noises in the entanglement are possible due to the surroundings of the subsystem under consideration.

## 3. MODEL AND RESULTS FOR FINITE CHAINS

Let us express the Hamiltonian of a quantum Ising chain as follows:

$$\mathcal{H} = -\sum_{i=1}^{L}\sigma_i^x - \lambda\sum_{i=1}^{L-1}\sigma_i^z\sigma_{i+1}^z, \qquad (6)$$

where $\lambda$ is the spin–spin interaction value and $\sigma_i^\alpha$ is the $\alpha$th component of the Pauli matrix at the $i$th lattice site.

For $\lambda = 1$, an infinitely long Ising chain exhibits a second-order phase transition, which is described by a conformal field theory (CFT) with a central charge of $c = 1/2$. However, we will start with finite-length chains.

Evidently, the simplest case corresponds to a dimer (two-qubit model). For $L = 2$, Hamiltonian (6) reduces to

$$\mathcal{H}_2 = -\sigma_1^x - \sigma_2^x - \lambda\sigma_1^z\sigma_2^z$$

$$= -\begin{pmatrix} \lambda & 1 & 1 & 0 \\ 1 & -\lambda & 0 & 1 \\ 1 & 0 & -\lambda & 1 \\ 0 & 1 & 1 & \lambda \end{pmatrix}. \qquad (7)$$

From this it follows that the energy levels of the quantum Ising dimer are as follows:

$$E_{1,2} = \pm\sqrt{\lambda^2+4}, \quad E_{3,4} = \pm\lambda. \qquad (8)$$

In both ferromagnetic ($\lambda > 0$) and antiferromagnetic ($\lambda < 0$) cases, the ground state (singlet) is nondegener-ate and possesses the same energy of $E_0 = -\sqrt{4 + \lambda^2}$. This ground state corresponds to the eigenvector

$$|\psi_0\rangle = \frac{1}{\sqrt{d}} \begin{pmatrix} \lambda + \sqrt{\lambda^2 + 4} \\ 2 \\ 2 \\ \lambda + \sqrt{\lambda^2 + 4} \end{pmatrix}, \quad (9)$$

where

$$d = 2[(\lambda + \sqrt{\lambda^2 + 4})^2 + 4]. \quad (10)$$

As is well known (see, e.g., [4, p. 32, 28, 36]), the two-qubit model with the wave function

$$|\psi\rangle = a_{1,1}|1,1\rangle + a_{1,-1}|1,-1\rangle \\ + a_{-1,1}|-1,1\rangle + a_{-1,-1}|-1,-1\rangle; \quad (11)$$

(normalized as $\langle\psi|\psi\rangle = 1$) is characterized by a concurrence of

$$C = 2|a_{1,1}a_{-1,-1} - a_{1,-1}a_{-1,1}|. \quad (12)$$

Applying these formulas to vector (9), we determine the following value of concurrence in the ground state of the Ising dimer [36, 37]:

$$C = [1 + (2/\lambda)^2]^{-1/2}. \quad (13)$$

Then, the entropy of entanglement of the two-qubit system is related to the concurrence as

$$S = -\frac{1 + \sqrt{1 - C^2}}{2} \ln\left(\frac{1 + \sqrt{1 - C^2}}{2}\right) \\ - \frac{1 - \sqrt{1 - C^2}}{2} \ln\left(\frac{1 - \sqrt{1 - C^2}}{2}\right), \quad (14)$$

and the magnitude of its fluctuations is [28]

$$\Delta S = C \ln\left[\frac{1}{C}(1 + \sqrt{1 - C^2})\right]. \quad (15)$$

Figure 1 shows the plots of $S(\lambda)$, $\Delta S(\lambda)$, and $\delta S(\lambda) = \Delta S/S$ for the quantum Ising dimer. According to these data, both the entanglement entropy $S$ and its fluctuation $\Delta S$ vanish at $\lambda = 0$ (i.e., for two uncoupled spins in a magnetic field). As the spin–spin interaction $\lambda$ increases, the entanglement entropy exhibits monotonic growth and tends to $\ln 2$ (residual entropy of a doubly degenerate, fully ordered state) in the limit as $\lambda \to \infty$. The entanglement entropy fluctuations exhibit a maximum near $\lambda = 1$ and then monotonically decrease until vanishing. The relative fluctuation $\delta S(\lambda)$ is a monotonically decreasing function (Fig. 1).

As can be seen from Fig. 1, the $S(\lambda)$ and $\Delta S(\lambda)$ curves intersect at $\lambda_f \approx 2.9447$. For $\lambda > \lambda_f$, the entanglement entropy fluctuations are relatively small ($\Delta S < S$). In contrast, for $0 < \lambda < \lambda_f$, the entanglement entropy exhibits strong fluctuations ($\Delta S \geq S$) and a simple notion of fluctuations as small deviations from equilibrium is inapplicable, so that it is necessary to take into account higher moments of the entanglement entropy and make allowance for the "interaction" of fluctuations.

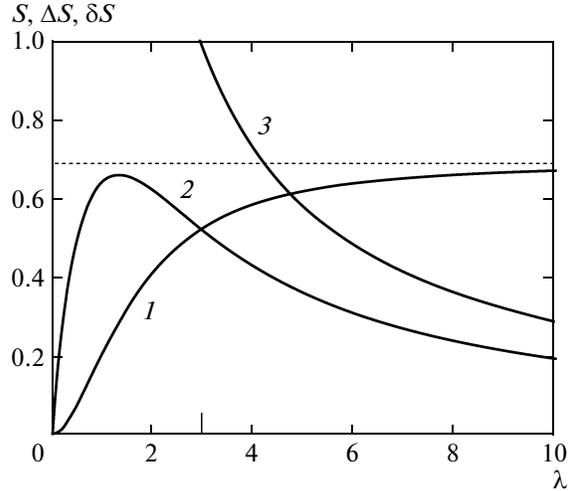

**Fig. 1.** Plots of (*1*) entanglement entropy $S$, (*2*) absolute fluctuation $\Delta S$, and (*3*) relative fluctuation $\delta S$ versus spin–spin interaction value $\lambda$ for two-qubit Ising system in the pure state with a minimum energy. Dashed line shows level of $S = \ln 2 \approx 0.693$; bar on abscissa axis corresponds to value of $\lambda_f = 2.9447...$.

The results of numerical calculations show that, as the chain length increases, the position of maximum in $\Delta S$ tends to $\lambda = 1$ (which is the point of the future quantum phase transition) and the height of this maximum slowly increases. The convergence implies that, far from $\lambda = 1$, the properties of finite long chains qualitatively correctly reflect the system behavior in the limit as $L = \infty$.

## 4. INFINITELY LONG CHAIN

The entropy of a semi-infinite part, $(-\infty, 0)$ or $(0, +\infty)$, of the linear Ising chain on the $(-\infty, +\infty)$ axis was calculated in [31, 32], where the reduced density matrix was found using a well-known method, whereby the quantum one-dimensional system was mapped to a classical two-dimensional (2D) model and this 2D model was calculated using the angle transfer matrices. According to this method, the reduced density matrix has the following Gibbs form:

$$\rho = \frac{1}{Z} e^{-\mathcal{H}'}, \quad (16)$$

where $Z = \text{Tr}[\exp(-\mathcal{H}')]$ and $\mathcal{H}'$ is the effective Hamiltonian of free fermions given by

$$\mathcal{H}' = \sum_{j=0}^{\infty} \varepsilon_j n_j. \quad (17)$$

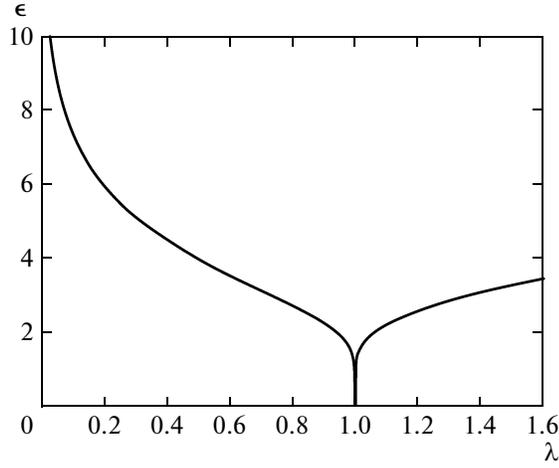

**Fig. 2.** Plot of distance $\epsilon$ between excitation levels as a function of $\lambda$.

Here, $n_j$ are the operators of occupation numbers (with unity or zero eigenvalues) and $\varepsilon_j$ are the excitation energies with equidistant spectrum

$$\varepsilon_j = \begin{cases} (2j+1)\epsilon, & \lambda < 1 \\ 2j\epsilon, & \lambda > 1. \end{cases} \quad (18)$$

where $j = 0, 1, 2, ...$ and $\epsilon$ is the distance between the energy levels. The latter quantity can be expressed as

$$\epsilon = \pi \frac{I(\sqrt{1-k^2})}{I(k)}, \quad (19)$$

where $k = \min\{\lambda, \lambda^{-1}\}$ and $I$ is the complete elliptic integral of the second kind as follows:

$$I(k) = \int_0^{\pi/2} \frac{d\theta}{\sqrt{1-k^2\sin^2\theta}}. \quad (20)$$

Figure 2 shows a plot of $\epsilon(\lambda)$, from which it is seen that, as $\lambda$ tends to zero or infinity, the distance between excitation levels increases (spectrum exhibits rarefaction). On the contrary, the distance between energy levels decreases in the vicinity of the critical point $\lambda_c = 1$, where the equidistant spectrum of excitations exhibits densification.

Using the above expressions for the reduced density matrix, one easily obtains the following formulas for the distribution function,

$$Z = \prod_{j=0}^{\infty}(1 + e^{-\varepsilon_j}) \quad (21)$$

and the entanglement entropy

$$S \equiv \langle \hat{S} \rangle = -\frac{\epsilon}{Z}\frac{\partial Z}{\partial \epsilon} + \ln Z. \quad (22)$$

Using these formulas, the entropy of entanglement of the quantum Ising chain in the ground state can be expressed in terms of infinite sums as [31]

$$S = \sum_{j=0}^{\infty} \frac{(2j+1)\epsilon}{1 + \exp[(2j+1)\epsilon]}$$
$$+ \sum_{j=0}^{\infty} \ln(1 + e^{-(2j+1)\epsilon}), \quad \lambda < 1 \quad (23)$$

for a disordered (paramagnetic) phase and as

$$S = \sum_{j=0}^{\infty} \frac{2j\epsilon}{1 + \exp(2j\epsilon)} + \sum_{j=0}^{\infty} \ln(1 + e^{-2j\epsilon}), \quad (24)$$
$$\lambda > 1$$

for the ordered (ferromagnetic) phase. In the right-hand side of Eq. (24), the term with $j = 0$ (making zero contribution) in the first sum is retained for the sake of unification of writing.

Using well known identities of the theory of elliptic functions [see Eqs. (A.1)–(A.4) in Appendix], Peschel [38] presented infinite sums in expressions (23) and (24) in a closed analytical form and obtained the following expressions for the entanglement (see also [39–41]):

$$S = \frac{1}{24}\left[\ln\left(\frac{16}{k^2 k'^2}\right) + \frac{4}{\pi}(k^2 - k'^2)I(k)I(k')\right], \quad (25)$$
$$\lambda < 1,$$

for the first branch and

$$S = \frac{1}{12}\left[\ln\left(\frac{k^2}{16k'}\right) \right.$$
$$\left. + \frac{4}{\pi}\left(1 - \frac{k^2}{2}\right)I(k)I(k')\right] + \ln 2, \quad \lambda > 1 \quad (26)$$

for the second branch ($k' = \sqrt{1-k^2}$ is the additional modulus).

Expressions (25) and (26) show that, in accordance with the CFT (see [42] and references therein), the entanglement diverges in the vicinity of the phase transition according to the following law [31, 32]:

$$S(\lambda \to 1) = \frac{c}{6}\ln\frac{1}{|1-\lambda|}, \quad (27)$$

where $c = 1/2$ is the central charge that determines the universality class of the (1+1)-dimensional Ising chain model.

Let us calculate the second moment of the entanglement entropy. By definition, we obtain

$$\langle(\hat{S})^2\rangle = \epsilon^2\frac{\partial^2 \ln Z}{\partial \epsilon^2} + \left(\epsilon\frac{\partial \ln Z}{\partial \epsilon} - \ln Z\right)^2. \quad (28)$$

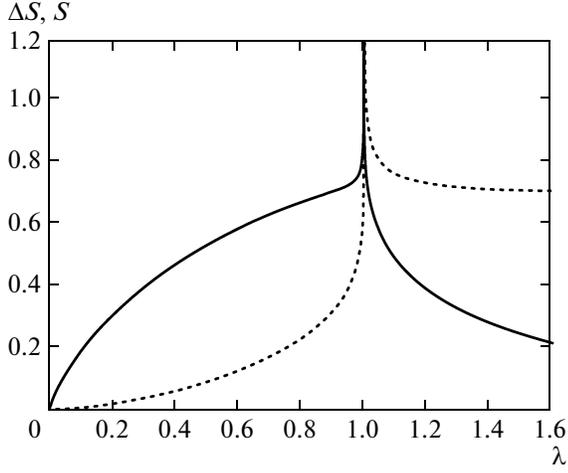

**Fig. 3.** Plot of entanglement entropy $S$ (dashed curve) and its fluctuations $\Delta S$ (solid curve) versus $\lambda$ for infinitely long Ising chain.

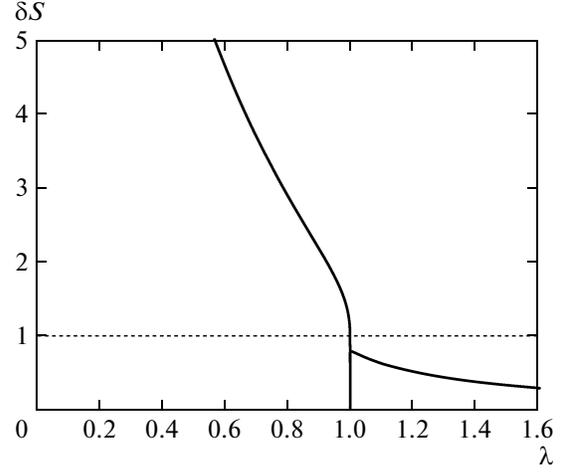

**Fig. 4.** Plot of relative fluctuation $\delta S = \Delta S/S$ of entanglement entropy versus $\lambda$ for infinitely long Ising chain. Dashed line indicates level of $\delta S = 1$.

Using expression (22), we obtain the following formula for the dispersion of the entropy:

$$D = \epsilon^2 \frac{\partial^2 \ln Z}{\partial \epsilon^2}. \qquad (29)$$

Taking the derivatives, we arrive at an expression for the fluctuations of entanglement in the form of an infinite series

$$\Delta S = \left[ \sum_{j=0}^{\infty} \frac{(\varepsilon_j/2)^2}{\cosh^2(\varepsilon_j/2)} \right]^{1/2}, \qquad (30)$$

where $\varepsilon_j$ is given by Eq. (18). Using Eqs. (A.5) and (A.6) (see Appendix), we eventually arrive at the following formulas for fluctuations of the entanglement entropy written in the closed analytical form

$$\Delta S = \frac{1}{\pi} I(k') \sqrt{\frac{2}{3} I(k)} \\ \times [k'^2 I(k) + (k^2 - k'^2) E(k)]^{1/2}, \quad \lambda < 1 \qquad (31)$$

$$\Delta S = \frac{2}{\pi} I(k') \sqrt{\frac{1}{3} I(k)} \\ \times \left[ \left(1 - \frac{k^2}{2}\right) E(k) - (1 - k^2) I(k) \right]^{1/2}, \quad \lambda > 1, \qquad (32)$$

where

$$E(k) = \int_0^{\pi/2} \sqrt{1 - k^2 \sin^2 \theta}\, d\theta \qquad (33)$$

is the complete elliptic integral of the second kind.

Equations (31) and (32) imply the following asymptotic behavior of the square of entanglement fluctuations in the vicinity of the phase transition:

$$(\Delta S)^2 = D \approx \frac{1}{12} \ln \frac{1}{|1-\lambda|}, \quad \lambda \longrightarrow 1. \qquad (34)$$

Thus, the dispersion of the entanglement entropy diverges at the point of transition by the same law (27) as the entanglement itself.

Formulas (27) and (34) yield the following scaling expansion for the second moment of the entropy operator near the phase transition point:

$$\langle (\hat{S}) \rangle = \left( \frac{1}{12} \ln \frac{1}{|1-\lambda|} \right)^2 + \frac{1}{12} \ln \frac{1}{|1-\lambda|} + \ldots \qquad (35)$$

## 5. DISCUSSION OF RESULTS

Figure 3 (solid curve) shows the behavior of fluctuations $\Delta S$ of the entanglement entropy as a function of $\lambda$ for an infinitely long Ising chain in the ground state. The dashed curve shows a plot of the $S(\lambda)$ function according to [31, 32]. The entanglement entropy of the infinite chain, as well as that of the finite chains (see Fig. 1), exhibits no fluctuations at $\lambda = 0$ (uncoupled spins) and at $\lambda = \infty$ (completely ordered doubly degenerate system). The entropy of the semi-infinite subsystem exhibits a singularity at the phase transition point. This phenomenon can be interpreted as the first-order phase transition conditioned by the presence of a reservoir, which takes place as a result of the averaging over states of the second semi-infinite subsystem.

It is interesting to consider the behavior of the relative fluctuation defined as $\delta S = \Delta S/S$. Figures 4 and 5 show the plots of $\delta S(\lambda)$ for an infinite Ising chain ($L = \infty$). In this system, as well as in finite chains, we have $\delta S \longrightarrow \infty$ on approaching the point of transition from the entangled to separable state. However, both the entanglement and its fluctuation vanish at the phase transition point (at $\lambda = 0$). The relative fluctuation also vanishes in the limit as $\lambda \longrightarrow \infty$.

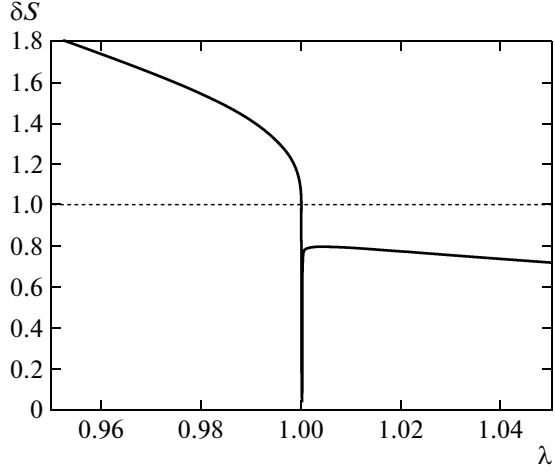

**Fig. 5.** Behavior of $\delta S(\lambda)$ in vicinity of phase transition point.

In a disordered phase (i.e., for $0 < \lambda < 1$), the function $\delta S(\lambda)$ monotonically decreases with increasing $\lambda$ and vanishes at $\lambda = 1$. During this, the entanglement fluctuations are greater than the entanglement proper ($\Delta S > S$) almost everywhere except for a narrow interval of $\lambda_f < \lambda < 1$ (where $\lambda_f = 0.999951...$) in which $\Delta S < S$.

For $\lambda > 1$, the curve of $\delta S(\lambda)$ exhibits a maximum at $\lambda_m \approx 1.0044$, where $\delta S_m \approx 0.7957$. Thus, in the ordered phase ($1 < \lambda < \infty$), for which the spontaneous magnetization $\langle \sigma^x \rangle \neq 0$, the relative entanglement fluctuation does not exceed 79.6% of the entanglement magnitude. This is the region of moderate and weak fluctuations.

Thus, in the limit of infinite length, the Ising chain exhibits a new effect, whereby the relative entanglement fluctuation vanishes at a finite value of the spin–spin interaction ($\lambda = 1$). This circumstance indicates that the entanglement fluctuations play a small role as compared to that of the entanglement itself near the point of the quantum phase transition in the system.

In recent years, the theory of quantum entanglement showed growing interest [39, 40] in using the concepts of the Tsallis entropy and Renyi entropy. The Renyi entropy, which was suggested in 1960 [43] as a generalization of the Shannon entropy, can be written in quantum-mechanical terms as

$$S_R^{(\alpha)} = \frac{1}{1-\alpha}\ln[\mathrm{Tr}(\rho^\alpha)]. \quad (36)$$

The Tsallis entropy was introduced in 1988 [44] as a generalization of the Boltzmann–Gibbs entropy and can be written in the quantum variant as

$$S_T^{(\alpha)} = \frac{1}{\alpha-1}[1 - \mathrm{Tr}(\rho^\alpha)]. \quad (37)$$

In the limit as $\alpha \longrightarrow 1$, both of these quantities convert into the von Neumann entropy given by Eq. (3). The Renyi and Tsallis entropies are interrelated as follows:

$$S_R^{(\alpha)} = \frac{1}{1-\alpha}\ln[1 + (1-\alpha)S_T^{(\alpha)}] \quad (38)$$

or

$$S_T^{(\alpha)} = \frac{1}{1-\alpha}[\exp((1-\alpha)S_R^{(\alpha)}) - 1]. \quad (39)$$

Let us apply the method of moments to the entropy operator. According to formula (37), the Tsallis entropy can be expanded as follows:

$$\begin{aligned}S_T^{(\alpha)} &= -\sum_{n=1}^{\infty}\frac{(\alpha-1)^{n-1}}{n!}\mathrm{Tr}(\rho\ln^n\rho)\\&= \langle\hat{S}\rangle - \frac{1}{2}(\alpha-1)\langle(\hat{S})^2\rangle + \ldots\end{aligned} \quad (40)$$

Thus, once the moments of the entropy operator are known, it is possible to reconstruct the Tsallis entropy and then find the Renyi entropy.

Calabrese and Cardy [31, 45] demonstrated that, by virtue of the conformal invariance, the moments of the reduced density matrix for an open quantum chain in the vicinity of the critical point can be expressed as follows:

$$\mathrm{Tr}\rho^\alpha \approx c_\alpha \xi^{-c(\alpha - 1/\alpha)/12}, \quad (41)$$

where $\xi$ is the correlation length (measured in units of the lattice constant) and $c_1 = 1$. Moreover, it can approximately be assumed that coefficients $c_\alpha$ are independent of $\alpha$ [46]. Taking into account that

$$\begin{aligned}\mathrm{Tr}\rho^\alpha &= \langle\exp[(1-\alpha)(-\ln\rho)]\rangle\\&= \sum_{n=0}^{\infty}\frac{(1-\alpha)^n}{n!}\langle(\hat{S})^n\rangle,\end{aligned} \quad (42)$$

we obtain the following relation:

$$\langle(\hat{S})^n\rangle = (-1)^n \frac{\partial^n}{\partial\alpha^n}\mathrm{Tr}\rho^\alpha\bigg|_{\alpha=1}. \quad (43)$$

Then, with allowance for Eq. (41), the entanglement entropy is expressed as

$$S = \langle\hat{S}\rangle \approx \frac{c}{6}\ln\xi \quad (44)$$

and the second moment of the entropy operator as

$$\langle(\hat{S})^2\rangle \approx \left(\frac{c}{6}\ln\xi\right)^2 + \frac{c}{6}\ln\xi. \quad (45)$$

Both of these expressions are consistent with formulas (27) and (35), respectively, since the quantum Ising chain has $\xi \approx 1/|1 - \lambda|$ and $c = 1/2$.

For a one-dimensional subsystem consisting of several long (on the order of $\xi$) parts spaced by large ($\gg \xi$) distances, the entanglement entropy exhibits the following asymptotic behavior [31, 45]:

$$S \approx \mathcal{A} \frac{c}{6} \ln \xi, \quad (46)$$

where $\mathcal{A}$ is the number of boundary (terminal) pints for all parts of the subsystem. This expression follows from a generalization of the asymptotic relation (41), which is written as

$$\mathrm{Tr}\rho^\alpha \approx c_\alpha \xi^{-\mathcal{A}c(\alpha - 1/\alpha)/12}. \quad (47)$$

According to formula (46), which is a one-dimensional analog of the aforementioned law of areas, the entanglement entropy of a multiply connected rarefied subsystem is additive, while the $S/\mathcal{A}$ ratio gives a normalized value (density) of the entanglement. For the entanglement fluctuations, formula (47) yields $\Delta S \sim \sqrt{\mathcal{A}}$ and, hence, $\delta S \sim 1/\sqrt{\mathcal{A}}$.

## 6. CONCLUSIONS

We have considered the behavior of fluctuations of the entanglement entropy for quantum Ising chains consisting of two parts connected at one point. It was established that, in a completely ordered or disordered state of the system, the absolute fluctuations of the entanglement vanish. However, on approaching the point of transition to the completely disordered state ($\lambda = 0$), the system passes through a region of strong fluctuations such that $\Delta S > S$, where the relative fluctuation of the entanglement entropy also exhibits unlimited growth.

For an infinitely long Ising chain and, accordingly, semi-infinite subsystems, the absolute fluctuation of the entanglement entropy diverges at the point of the quantum phase transition. Upon approaching this critical point, the square of the magnitude of fluctuations tends to infinity by the logarithmic law, while the amplitude of this singularity is controlled by the CFT. The ratio of the square of fluctuations to the entropy at the critical point is unity.

The region of strong fluctuations accounts for 99.995% of the total domain of disordered states of the system on the $\lambda$ axis. It is only in the immediate vicinity of the critical point that the relative fluctuation decreases and completely vanishes at the phase transition point ($\lambda_c = 1$). On the contrary, the entanglement fluctuations in the ordered phase are always smaller than the entanglement proper. This entire domain is characterized by relatively weak fluctuations.

ACKNOWLEDGMENTS

The author is grateful to participants of the Seminar of the Spin Dynamics and Spin Computing Laboratory of the Theoretical Department of the Institute of Problems of Chemical Physics (Chernogolovka) for fruitful discussion and to I. Peschel for useful communications via e-mail.

This study was supported by the Presidium of the Russian Academy of Sciences (program nos. 18 and 21).

## APPENDIX

Below we will obtain identities that were used to transform the infinite sums in expressions for the entanglement entropy and its fluctuations. Let us proceed from the following identities proved in the theory of elliptic functions ([47, Ch. 21, Example 10] and [48, Eqs. (16.37.2) and (16.37.3)]):

$$\prod_{j=0}^{\infty}(1 + q^{2j+1}) = \left(\frac{16q}{k^2 k'^2}\right)^{1/24} \quad (A.1)$$

and

$$\prod_{j=0}^{\infty}(1 + q^{2j}) = 2\left(\frac{k^2}{16qk'}\right)^{1/12}, \quad (A.2)$$

where $q = \exp(-\pi I(k')/I(k))$ is Jakobi parameter.

Differentiating logarithms of Eqs. (A.1) and (A.2) with respect to $k$ and accomplishing transformations, we obtain the following identities:

$$\sum_{j=0}^{\infty}(2j+1)\frac{q^{2j+1}}{1+q^{2j+1}} = \frac{1}{24}\left[1 - (1-2k^2)\left(\frac{2I}{\pi}\right)^2\right] \quad (A.3)$$

and

$$\sum_{j=0}^{\infty} 2j\frac{q^{2j}}{1+q^{2j}} = \frac{1}{12}\left[\left(1 - \frac{k^2}{2}\right)\left(\frac{2I}{\pi}\right)^2 - 1\right], \quad (A.4)$$

which also follow from [47, Eqs. (16.23.11) and (16.23.12)].

Also differentiating Eqs. (A.3) and (A.4) with respect to $k$, we eventually arrive at the other two identities:

$$\sum_{j=0}^{\infty}(2j+1)^2 \frac{q^{2j+1}}{(1+q^{2j+1})^2} = \frac{2I^3}{3\pi^4}[(1-k^2)I - (1-2k^2)E] \quad (A.5)$$

and

$$\sum_{j=0}^{\infty}(2j)^2 \frac{q^{2j}}{(1+q^{2j})^2} \qquad (A.6)$$
$$= \frac{4I^3}{3\pi^4}\left[\left(1-\frac{k^2}{2}\right)E - (1-k^2)I\right].$$

In the above equations, $I = I(k)$ and $E = E(k)$ are the complete elliptic integrals. In deriving Eqs. (A.3)–(A.6), we used the following useful relations:

$$\frac{dq}{dk} = \frac{\pi^2 q}{2kk'^2 I^2},$$
$$\frac{dI}{dk} = \frac{1}{k}\left(\frac{E}{k'^2} - I\right),$$

and

$$IE' + I'E - II' = \frac{\pi}{2},$$

where $I' = I(k')$ and $E' = E(k')$.